\documentclass{article}
\usepackage{icrctc07}

\newcommand{\gray}[1]{$\gamma$-ray{#1}}

\newcommand{\pubjournal}[6] {#1, #2 {\bf #3}, #4 (#5).}

\newcommand{\apj}{{\it ApJ}}
\newcommand{\sci}{{\it Science}}

\newcommand{\nimA}{{\it Nuc. Instr. Meth. Phys. Res. A}}
\newcommand{\jgr}{{\it J.\ Geophys.\ Res.}}
\newcommand{\jgpr}{{\it J.\ Geo.\ Phys.\ Res.}}

\title{Gamma-ray albedo of the moon}
\shorttitle{Gamma-ray albedo of the moon}

\abstract{We use the GEANT4 Monte Carlo framework to calculate the \gray{} 
albedo of the Moon due to interactions of cosmic ray (CR) 
nuclei with moon rock.
Our calculation of the albedo spectrum agrees with the EGRET data. 
We show that
the spectrum of \gray{s} from the Moon 
is very steep with an effective cutoff around 4 GeV (600 MeV
for the inner part of the Moon disc).
Apart from other astrophysical sources,
the albedo spectrum of the Moon is well understood, including
its absolute normalisation; this makes it a useful
``standard candle'' for \gray{} telescopes,
such as the forthcoming Gamma Ray Large Area Space Telescope (GLAST).
The albedo flux depends on the incident CR spectrum which
changes over the solar cycle.
Therefore, it is possible
to monitor the CR spectrum using the albedo \gray{} flux.
Simultaneous measurements of CR
proton and helium spectra by the Payload for Antimatter 
Matter Exploration and Light-nuclei Astrophysics (PAMELA), 
and observations of the albedo 
\gray{s} by the GLAST Large Area Telescope (LAT), can be used to test the model
predictions and will enable the GLAST LAT to monitor the
CR spectrum near the Earth beyond the lifetime of PAMELA.
}

\begin{document}

\authors{I. V. Moskalenko$^{1,2}$, T. A. Porter$^{3}$ }
\shortauthors{Moskalenko and Strong}
\afiliations{
$^1$Hansen Experimental Physics Laboratory, Stanford University, Stanford, CA 94305, U.S.A.\\ 
$^2$Kavli Institute for Particle Astrophysics and Cosmology, Stanford University, CA 94309, U.S.A.\\
$^3$Santa Cruz Institute for Particle Physics, University of California, Santa Cruz, CA 95064, U.S.A.}
\email{imos@stanford.edu}

\maketitle

\section{Introduction}
Interactions of Galactic CR nuclei with the atmospheres of the Earth and 
the Sun produce albedo \gray{s} due to the
decay of secondary neutral pions and kaons (e.g., \cite{Seckel1991,Orlando2007}).
Similarly, the Moon emits \gray{s} due to CR interactions
with its surface \cite{Morris1984,Thompson1997}; low energy \gray{} 
spectroscopy data acquired by the Lunar Prospector were used to map
the elemental composition of the Moon surface \cite{Lawrence1998,Prettyman2006}.
However, contrary to the CR interaction with the gaseous
atmospheres of the Earth and the Sun, the Moon surface is solid, 
consisting of rock, making its albedo spectrum unique.

Due to the kinematics of the collision, the secondary particle cascade 
from CR particles hitting the Moon surface at small zenith angles
develops deep into the rock making it difficult for \gray{s} to get out.
A small fraction of all produced pions, splash albedo pions,
are mostly low energy ones thus producing the soft spectrum \gray{s}. 
High-energy \gray{s} can be
produced by CR particles hitting the Moon surface in a close-to 
tangential direction.
However, since it is a solid target, only the very thin
limb contributes to the high energy emission.

The \gray{} albedo of the Moon has been calculated by Morris
\cite{Morris1984} using a Monte Carlo code 
for cascade development in the Earth's atmosphere that was modified 
for the Moon conditions.
However, the CR spectra used as input in \cite{Morris1984} differ considerably
from recent measurements by AMS and BESS. Besides,
due to the lack of accelerator data and models a number 
of approximations and ad-hoc assumptions were required to calculate
the hadronic cascade development in the solid target.

The Moon has been detected by the EGRET as a point source 
with integral flux $F(>$$100\ {\rm MeV})=(4.7\pm0.7)\times10^{-7}$ cm$^{-2}$
s$^{-1}$ \cite{Thompson1997}, $\sim$24\% below the predictions
\cite{Morris1984}.
The observed spectrum is steep and yields only the upper limit 
$\sim$$5.7\times10^{-12}$ cm$^{-2}$ s$^{-1}$ above 1 GeV.

We report preliminary results for 
calculations of the \gray{} albedo from the Moon
using the GEANT4 \cite{Agostinelli2003}
framework code and discuss the consequences of 
its measurement by the upcoming GLAST mission.

\section{Monte Carlo simulations}

\begin{figure}[t]
\begin{center}
\includegraphics[width=2.8in]{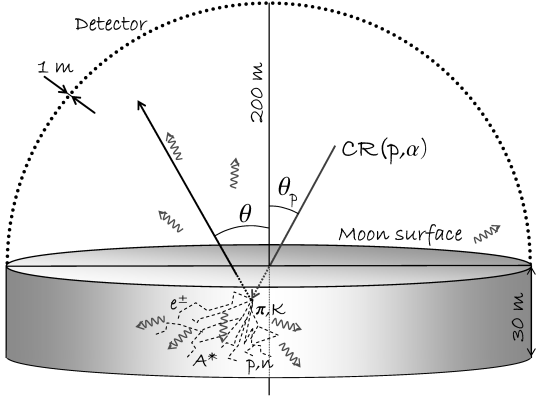}
\end{center}
\caption{Beam/target/detector setup for simulating CR interactions in moon 
rock. 
The primary beam enters the moon rock target with incident polar 
angle $\theta_p$. 
Secondary \gray{s} are emitted with polar angle $\theta$. 
The detection volume surrounds the target.}
\label{fig1}
\end{figure}

In the present work, we use version 8.2.0 of the GEANT4 
toolkit.
Figure~\ref{fig1} illustrates our beam/target/detector setup for 
simulating CR interactions in the Moon.
The primary CR beam (protons, helium nuclei) is injected at different
incident angles into a moon rock target.
We take the composition of the moon rock to be 
45\% SiO$_2$, 22\% FeO, 11\% CaO, 10\% Al$_2$O$_3$, 
9\% MgO, and 3\% TiO$_2$ by weight, consistent with mare basalt 
meteorites and Apollo 12 and 15 basalts 
\cite{Lawrence1998,Anand2003,Prettyman2006}.
A thin hemispherical detector volume surrounding the target is used to 
record the secondary \gray{} angular and energy distributions in the simulation.

\begin{figure}[t]
\begin{center}
\includegraphics[width=2.8in]{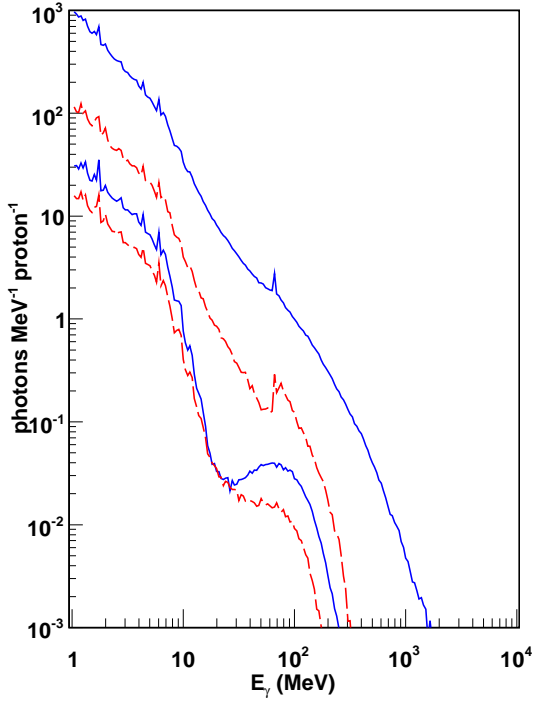}
\end{center}
\caption{\gray{} yield per proton interaction integrated over
all emission angles from the Moon surface.
Line-styles: red-dashed, $\cos\theta_p = 1$; 
blue-solid, $\cos\theta_p = 0.1$. 
Line-sets: lower, $E_p = 500$ MeV; upper, $E_p = 10000$ MeV.}
\label{fig2}
\end{figure}

The \gray{} yield $dY_\gamma(E_p,\cos\theta_p)/dE_\gamma d\cos\theta$
is calculated using the GEANT4 beam/target setup with
a Monte Carlo method; here $E_p$ is the kinetic energy per nucleon of 
the incident particle, $\theta_p$ is the incident polar angle,
$E_\gamma$ is the energy of \gray{s}, and $\theta$ is the 
polar photon emission angle.

Figure~\ref{fig2} shows the secondary \gray{} yields integrated over 
all emission angles outward from the Moon surface for
protons with $E_p = 500$ MeV and 5000 MeV at incident 
angles $\cos\theta_p = 0.1$ and 1, respectively.
The shapes of the yield curves for different incident angles are very similar
to each other for the case of low energy protons where the secondary particles
(pions, kaons, neutrons, etc.) are produced nearly at rest. 
In this case
the \gray\ emission is produced in a number of processes:
pion and kaon decay, secondary electron and positron bremsstrahlung, 
and so forth. 
A considerable flux of \gray{s} is produced in nuclear 
reactions such as neutron capture and nonelastic scattering
\cite{Lawrence1998,Prettyman2006};
the features below $\sim$10 MeV are due to nuclear de-excitation lines,
where the most prominent contribution comes from oxygen nuclei.
In the high energy case,
the secondary distribution for protons incident near zenith has a 
cutoff above $\sim$500 MeV.
Further away from zenith higher yields of secondary \gray{s} are 
produced while the spectrum of \gray{s} becomes
progressively harder.
This is a result of the cascade developing mostly in the forward direction:
for near zenith primaries, most high energy secondary \gray{s} will be 
absorbed in the target, while a small fraction of produced pions and kaons,
splash albedo particles, mostly low energy ones, produce the 
soft spectrum \gray{s}; further from zenith, the high energy secondary
\gray{s} will shower out of the Moon surface.

\begin{table*}[tbh]
\begin{tabular}{|l||c|c|c|c|c|c|c|c|c|c|}
\hline
particle & $J_0$ & 
$a_1$ & $b_1$ & $c_1$ & 
$a_2$ & $b_2$ & $c_2$ &
$a_3$ & $b_3$ & $c_3$ \\
\hline\hline
proton & 
$1.6\times10^4$ &
 1       & 0.458 & 2.75    &
--3.567 & 0.936 & 4.90 &
 $4.777\times10^5$ & 14.4   & 6.88 \\
helium &
$1.6\times10^3$ &
 1       & 1.116 & 3.75  &
2.611  & 4.325 & 3.611  &
0.219 & 0.923 & 2.58 \\
\hline
\end{tabular}
\footnotetext{}{\footnotesize
Units of the flux: m$^{-2}$ s$^{-1}$ sr$^{-1}$ (GeV/nucleon)$^{-1}$.}
\caption{Fits to local interstellar CR spectra}
\label{Table1}
\end{table*}

\section{Calculations}
To calculate the Moon albedo at an arbitrary solar modulation level, 
we use the local interstellar spectra (LIS) of CR protons and helium as fitted to the 
numerical results of GALPROP propagation model 
(reacceleration and plain diffusion models, Table 1 in \cite{Ptuskin2006});
the CR particle flux at an arbitrary phase of solar activity at 1 AU
can then be estimated using the force-field approximation \cite{Gleeson1968}.

To fit the LIS CR spectra we choose a function of the form:
\begin{equation}
\frac{dJ_p}{dE_k}=J_0 \sum_{i=1}^3 a_i (E_k+b_i)^{-c_i},\
\end{equation}
where the flux units are m$^{-2}$ s$^{-1}$ sr$^{-1}$ (GeV/nucleon)$^{-1}$
and the parameter values are given in Table~\ref{Table1}.
The latter are not unique and other sets could produce similar
quality fits, but this does not affect the final results.

\begin{figure}[t]
\begin{center}
\includegraphics[width=2.8in]{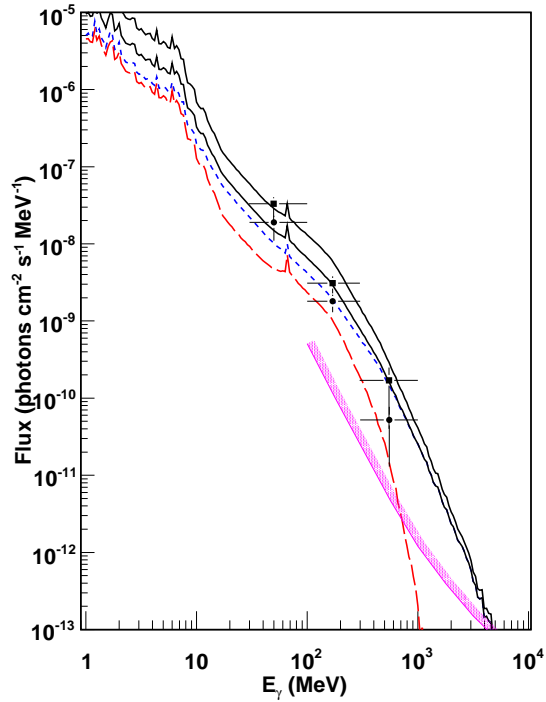}
\end{center}
\caption{Calculated \gray{} albedo spectrum of the Moon.
Line-styles: black-solid, total; blue-dotted, limb -- outer $5'$; 
red-dashed, centre -- inner $20'$.
Upper solid line: $\Phi_0 = 500$ MV; lower solid line: $\Phi_0 = 1500$ MV.
Data points from the EGRET \cite{Thompson1997} 
with upper and lower symbols corresponding 
to periods of lower and higher solar activity, respectively.
The differential 1 year sensitivity of the 
LAT is shown as the hatched region.}
\label{fig3}
\end{figure}

Figure \ref{fig3} shows the calculated total \gray{} albedo spectrum 
for CR protons and helium compared to the EGRET data for periods of lower solar 
(upper solid: $\Phi = 500$ MV) and higher solar activity 
(lower solid: $\Phi = 1500$ MV). Taking into 
account that the exact CR spectra during the EGRET observations
are unknown, the agreement with the data is remarkable.
The broken lines show the spectra
from the limb (outer $5'$) and the central part of the 
disc ($20'$ across) for the case of higher solar activity.
As expected, the spectra from the limb and the
central part are similar at lower energies ($<$10 MeV); at high
energies the central part exhibits a softer spectrum so that virtually
all photons above $\sim$600 MeV are emitted by the limb.

The clear narrow line at 67.5 MeV in Figure~\ref{fig3} 
is due to neutral pions decaying near rest.
The line perhaps always exists due to the splash albedo particles,
but in case of a thin and/or gaseous target (usual in astrophysics)
it is indistinguishable from the background \gray{s}.
For the case of a thick solid rock target, and thus fast energy 
dissipation of the particle cascade, and because of the 
kinematics of the interaction, the continuum \gray\ background is much 
smaller revealing the narrow line from pion decay at rest.

\section{Discussion and Conclusions}

The GLAST LAT is scheduled for launch by NASA in winter of 2008. 
It will have 
superior angular resolution and effective area, and its field of view (FOV)
will far exceed that of its predecessor EGRET \cite{McEnery2004}.
The LAT will scan the sky continuously providing complete sky coverage
every two orbits ($\sim$3 hr). 
About 20\% of the time the Moon 
will be in the FOV at different viewing angles.
The point spread function (PSF) of the instrument is
$\sim$0.8$^\circ$, at 1 GeV, but
reduces dramatically at higher energies: $\sim$0.5$^\circ$ at 2 GeV
and $\sim$0.2$^\circ$ at 10 GeV. 

The Moon with its steep albedo spectrum
presents almost a black spot on the \gray{} sky 
above $\sim$4 GeV. 
The central part of the Moon has an even steeper 
spectrum with a cutoff at $\sim$600 MeV.
The albedo spectrum of the Moon is well understood while 
the Moon itself is a ``moving target'' passing through high Galactic 
latitudes and the Galactic centre region. 
This makes it a useful
``standard candle'' for the GLAST LAT at energies below 1 GeV.  
A simultaneous presence of the PAMELA on-orbit 
capable of measuring protons and light nuclei with
high precision provides a necessary input for accurate
prediction of the albedo flux and a possible independent 
calibration of the GLAST LAT.
An additional bonus of such a calibration is the possibility to use
GLAST observations of the Moon to monitor the CR spectra
near the Earth beyond the projected 3 yr lifetime of the PAMELA.

The line feature around 67.5 MeV  from $\pi^0$-decay 
produced by CR particles in the solid rock target is interesting.
The lower energy limit of the LAT instrument
is below 20 MeV while the energy resolution is $\sim$15\% at 100 MeV, and 
improves at higher energies.
With a suitable event selection it may be possible to observe the line. 
There is no other astrophysical object predicted to
produce such a narrow line and there is no other line expected
except, perhaps, from dark matter annihilation.
A possibility for energy calibration at higher energy is provided
by the steep albedo spectrum above 100 MeV:
a small error in the energy determination will result in a large
error in the intensity.


\section{Acknowledgements}
We thank Bill Atwood, Seth Digel, Robert Johnson, 
and Denis Wright for many useful discussions.
I.\ V.\ M.\ acknowledges partial support from NASA APRA grant.
T.\ A.\ P.\ acknowledges partial support from the US Department of Energy.

\end{document}